# Aadhaar-Based Unified Payment Solution

Sankalp Bagaria
Email : sankalp [dot] nitt [at] gmail [dot] com

**Abstract:** In this paper, we propose to build an Aadhaar-Based Unified Payment Solution. The key idea is that a virtual wallet will be linked to the Aadhaar card number of the customer. After that, any identification unique to the person and linked with the Aadhaar card, be it something the person knows like secret Internet-banking password, be it something s/he carries like debit card/ credit card, something s/he owns like fingerprints, voice, email-id or somewhere s/he is like house or office address, can be used for money transfer from the sender's Aadhaar card–linked virtual wallet to the receiver's Aadhaar card–linked virtual wallet, whose any unique ID is known to the sender. If the sender knows the receiver's email-id, s/he can transfer money to his/ her Aadhaar card–linked virtual wallet using the email-id. And, if the sender knows receiver's mobile number but not email-id, s/he can use the mobile number to transfer the money to his/ her Aadhaar card–linked virtual wallet. And so on.

**Keywords:** Aadhaar card, Unified Payment Solution, Integrated Payment Solution, Biometric Payment, Voice based Payment, Mobile based Payment, e–Payment, Payment Solution, e-Commerce

**Introduction:** There are various solutions available in the market such as Pay Pal which integrates the email–id with the bank account, Paytm which integrates mobile number with bank account etc. Almost all banks provide facilities like Internet-banking and phone–banking. Most banks also allow statements to be delivered to the customer's email–id and/ or house or office address as requested by the customer. ICICI is going to be the first bank in the country to roll out voice authentication over phone number. Government of India is making great efforts on building a uniform payment system linking Aadhaar card, email–id, phone number and bank account. This paper wishes to add to that effort.

This solution unifies all prevalent mechanisms of money transfer by allowing transfer of the money from any Aadhaar–linked virtual wallet using any of the sender's ID is to the receiver's Aadhaar–linked virtual wallet using any of the receiver's ID. The advantage of this solution over existing solutions is – we can transfer money if we know any of the many IDs of the receiver. The customer would build an Aadhaar–linked virtual wallet, register all his/ her ID like phone number, email–id, home address, office address, biometrics etc and add some money to it from his/ her bank account. Those, who do not have bank account, can go to designated outlets to have money added to their Aadhaar card–linked virtual wallet by paying cash. Then this wallet can be used to transfer money from/ to it to/ from other Aadhaar–linked wallets.

It also proposes to build a web application and a mobile app to link the email–id and phone numbers stored in the customer's mobile with his/ her contacts' respective Aadhaar card–based virtual wallets. It is done by searching and matching the email–id and phone number against their Aadhaar card number stored in the Aadhaar–wallet database. Then the customer will be able to use his mobile app or web application to not only shop online like in Paytm but also send/ receive money from/ to his/her wallet to/ from friends/ relatives' Aadhaar card–linked virtual wallet using any ID known to her/ him as explained above. Fingerprints, retina scan or any other ID can be used to do transactions.

**Related Work:** Pay Pal [1] uses email–id and password for shopping and doing transactions. Paytm [2] allows shopping using mobile number. ICICI [3] is to roll out voice authentication from registered mobile number for allowing transactions over phone. Axis Bank [4] is launching Ping Pay, which will allow transactions using Facebook, Whatsapp, Twitter, SMS or email. [5] lists 8 companies that are working on authentication using biometrics. [6] talks about linking Aadhaar card, email–id, phone number and bank account. [7], [8] and [9] are the current efforts in Aadhaar-based Unified Payment System by the Government of India.

**Unified Payment Solution (UPS):** As mentioned in the abstract, the key idea is to enable the transaction using any unique ID of a person like his/her email–id, mobile number, voice, fingerprints, bank account, home/ office address etc from his/ her Aadhaar–linked wallet to another person's Aadhaar–linked wallet using any of receiver's (same as that of sender or different from it) ID known to the sender.

An individual may use his/ her email–id or phone number to send money by using the web application or mobile app which has been linked with the Aadhaar card–based virtual wallets beforehand by matching the email–ids and phone numbers with Aadhaar card number from the Aadhaar–wallet database. The customer can request transfer of money by calling the app centre directly also. The app centre, on receiving the call, will authenticate the transaction after verifying his/ her voice using voice biometrics. Another option is to allow the sender to use his/ her registered email–id or phone number to send an email or sms to the server citing the amount to be transferred and the email–id or phone number or any other ID of the receiver to whom the transfer is to be made.

The idea is that a sender may ask mobile number of the owner of the shop, buy products and pay for them by transferring the money to the shopkeeper's Aadhaar–linked virtual wallet by using mobile app. The customer would enter the mobile number of the shopkeeper, confirm it and enter the amount to be transferred. The server matches mobile number with the Aadhaar number of the shopkeeper and transfers the amount to his/ her Aadhaar–linked wallet from the customer's wallet.

A facility like sending money–order to family in a remote village by calling the Aadhaar–wallet centre is to be included. Transactions using other ID like fingerprint biometrics; retina scan etc at designated PoS is also to be included. The transactions can be made directly to the shopkeeper's wallet through registered PoS if the PoS s linked to the shopkeeper's Aadhaar-based wallet. The standard method of swiping credit card/ debit card at PoS is also available.

**Use Cases:**

**Home address to Phone number:** A person in a remote village sends money–order to Aadhaar-based wallet centre citing phone number to which the money has to be credited. Aadhaar-linked wallet centre identifies the Aadhaar card number to which the phone number belongs to. Then the server credits the amount to the Aadhaar–linked virtual wallet to which the phone number, mentioned in the money order, belongs to and sends the receipt to the village customer.

**Fingerprint to Phone number**: A person swipes his/ her fingerprint on designated PoS (Point–of–Sale) and punches the phone number of the person to whose wallet money is to be transferred. The PoS is linked to a centralised server that matches the customer's fingerprint to the fingerprints stored in the Aadhaar–database and locates his/ her Aadhaar number. On finding a match, the server deducts the money from the Aadhaar–linked wallet of the sender and credits the amount to the Aadhaar–linked wallet whose phone number has been fed by the sender.

If the Biometric PoS is registered in the shopkeeper's name, phone number need not be given. The money entered at PoS will be deposited directly to the Aadhaar–linked virtual wallet, linked with the PoS.

**Email–id to another email-id:** The customer sends email to the wallet server from his/ her registered email–id, citing email–id of the receiver and the amount to be transferred. The server finds the Aadhaar number corresponding to both the email–ids from its database. Then it debits the customer's wallet by the amount requested and credits the receiver's Aadhaar–linked wallet.

**Phone number to phone number:** On his/ her mobile app, the customer clicks against the name of the receiver (which has been matched to his/ her phone number) to whom the sender wants to transfer the money and then enters the amount. (The phone numbers of the sender's contacts have been synch-ed with their Aadhaar number–based virtual wallets beforehand by the Aadhaar server at the time of installing the mobile app.) The server authenticates the transaction, debits the customer's Aadhaar–linked virtual wallet and credits the receiver's Aadhaar–linked virtual wallet.

**Voice Biometrics to Email–id:** The customer calls the Aadhaar-wallet centre and asks to transfer money from his/ her wallet to the email–id of the receiver. The server does voice biometrics, validates the customer's voice and finds sender's Aadhaar card number from its database. It also parses the voice sample and interprets the information ie, the email-id and the amount to be transferred. The server then finds the Aadhaar card number related to the email-id and authenticates the transaction from the sender's wallet to the receiver's wallet.

**Voice Biometrics to Home address**: A factory worker may call Aadhaar–wallet centre and give his/ her Aadhaar number and asks for transfer of money from his/ her wallet to his/ her family at his/ her village home's address. The app centre will authenticate his/ her voice and send the money–order to his/ her home address at the village.

**Debit Card to Aadhaar–Wallet:** A person may use a debit card linked to his/ her Aadhaar card–based virtual wallet. The Point–of–Sale is linked with the shopkeeper's wallet. When the person swipes his/ her card at the shopkeeper's PoS, the entered amount is transferred from the card–holder's wallet to the shopkeeper's wallet.

**Secret Password to Email–id:** The customer logs in to his/ her web application and uses his/ her secret password to access his/ her Aadhaar-linked virtual wallet and transfer money to the email–id of one of his/ her contacts. Email-ids of the contacts had been linked with their respective Aadhaar card–based virtual wallets at the time of the installation of the application or when s/he synch-ed it with the database of the Aadhaar-wallet.

**Aadhaar Wallet from /to Bank Account/ Cash:** The sender can approach designated outlets to deposit money to his/ her virtual wallets by cash or by using a bank–issued credit card/ debit card. It can be done through Internet-banking also. S/he can also request cash or transfer of money to his/ her bank account from his/ her own virtual wallet at the designated outlets. S/he can do it online too.

**Conclusion:** This method will help the users to do the transactions between Aadhaar–linked virtual wallets using email–id, mobile number or any other ID of the receiver. This product will not only make shopping easier but also enable easy transfer of money between people. It will also remove the need to carry the debit card/ credit card always. A multi–factor authentication can be used for high value transactions. An appropriate ceiling on the amount of daily transactions can be put to prevent losses due to theft. The receiver can transfer his/ her money to bank account from his/ her UPS wallet at any convenient time. Those, who do not have bank accounts, can deposit/ receive cash to/ from their wallets at designated service centres.